\documentclass[pra,twocolumn,showkeys,showpacs,superscriptaddress,amsmath,amsfonts]{revtex4-1}
\usepackage[utf8x]{inputenc}
\usepackage{microtype}
\usepackage{lmodern}
\usepackage{graphicx}
\usepackage{hyperref}

\newcommand{\bra}[1]{\langle#1\rvert}
\newcommand{\ket}[1]{\lvert#1\rangle}
\newcommand{\mean}[1]{\langle#1\rangle}
\allowdisplaybreaks[1]

\begin{document}

\title{Green's function variational approximation: significance of physical constraints}

\author{Krzysztof Bieniasz}
\email{krzysztof.bieniasz@uj.edu.pl}
\affiliation{Marian Smoluchowski Institute of Physics, Jagiellonian University,
             {\L}ojasiewicza 11, PL-30348 Krak\'ow, Poland}
\affiliation{Department of Physics and Astronomy, University of British Columbia,
             Vancouver, British Columbia, Canada V6T 1Z1}
\affiliation{\mbox{Quantum Matter Institute, University of British Columbia,
             Vancouver, British Columbia, Canada V6T 1Z4}}

\author{Mona Berciu}
\affiliation{Department of Physics and Astronomy, University of British Columbia,
             Vancouver, British Columbia, Canada V6T 1Z1}
\affiliation{\mbox{Quantum Matter Institute, University of British Columbia,
             Vancouver, British Columbia, Canada V6T 1Z4}}

\author{Andrzej M. Ole\'s}
\affiliation{Marian Smoluchowski Institute of Physics, Jagiellonian University,
             {\L}ojasiewicza 11, PL-30348 Krak\'ow, Poland}
\affiliation{Max-Planck-Institut f\"ur Festk\"orperforschung,
             Heisenbergstra{\ss}e 1, D-70569 Stuttgart, Germany}

\date{\today}

\begin{abstract}
We present a calculation of the spectral properties of a single charge 
doped at a Cu($3d$) site of the Cu-F plane in KCuF$_{3}$.
The problem is treated by generating the equations of motion for the 
Green's function by means of subsequent Dyson expansions and solving the 
resulting set of equations. This method, dubbed the variational 
approximation (VA), is both very dependable and flexible, since it is 
a systematic expansion with precise control over elementary physical 
processes. It allows for deep insight into the underlying physics of 
polaron formation as well as for inclusion of many physical 
constraints, such as excluding crossing diagrams and double occupation 
constraint, which are not included in
the Self-Consistent Born Approximation (SCBA). Here we examine the role 
and importance of such constraints by analyzing various spectral functions
obtained in VA and in SCBA.
\end{abstract}


\maketitle

\section{Introduction}
\label{sec:intro}

Strongly correlated electron systems with long range ordered ground 
states exhibit a variety of interesting and complicated phenomena  
\cite{Kha05,Rei05,Wro10,Brz15}. Among the principle problems of interest 
are those of itinerant charge propagation and its coupling to the 
polarized background, \emph{e.g.} a hole propagating in an 
antiferromagnetic CuO$_{2}$ plane of a high $T_{c}$ cuprate superconductor 
\cite{Mar91}. Such a system is described by the well known $t$-$J$ model 
with a SU(2) symmetric Heisenberg Hamiltonian, where fluctuations play a 
crucial role in the coherent propagation of the charge. In particular, 
because of the total spin conservation, in the absence of fluctuations the 
only avenue of coherent charge propagation would be through Trugman loops, 
a~self-healing effective hopping \cite{Tru88}.

On the other hand, the orbital models offer a much wider range of 
charge propagation scenarios. In a system with long range orbital 
order, the exchange interaction always has a lower symmetry than SU(2), 
since the fluctuations are suppressed, while the orbital flavor might 
not necessarily be conserved. Because of that, even if orbital 
fluctuations are neglected, the charge propagation itself can lead to 
orbital (de)excitations in the system \cite{vdB00}.

Smaller fluctuations of the orbital exchange models mean that orbital 
systems behave more classically than their spin analogues. For instance, 
$t_{2g}$ orbital systems exhibit Ising exchange \cite{Dag08}, 
\emph{i.e.}, no fluctuations at all. However, weak hole propagation is 
still allowed because of three-site processes, which are of the same 
energy scale as the regular exchange itself. On the other hand, $e_{g}$ 
orbital systems, such as KCuF$_{3}$, are more complicated, with 
fluctuations only slightly suppressed due to the orbital symmetries, 
accompanied by very strong orbital non-conservation, which is kinetic 
in nature and governed by the hopping energy scale~$t$ \cite{Fei05}. 
Since this is the dominating interaction of the model, allowing for 
coherent propagation even if exchange fluctuations and three site terms 
are neglected, one would expect a~quasiparticle (QP) dispersion on the 
scale of the hopping energy $t$. The fact that previous research, based 
on the popular self-consistent Born approximation (SCBA) method, 
suggested extremely small QP dispersion seams to contradict this 
intuition. Hence it was suggested that the numerous simplifications 
required by the SCBA are in fact too restrictive and a more resilient 
approach is needed to better understand the effective models of 
$e_{g}$ systems. Here we make such an attempt via the variational 
approximation (VA) \cite{Ber11,Tro13,Ebr14}, or generating equations of 
motion for the Green's function.

\section{Methodology}
\label{sec:method}

An effective two-dimensional (2D) orbital model of a KCuF$_{3}$ 
ferromagnetic plane can be derived by second order canonical 
perturbation expansion 
using $\sigma$-bond hopping $t$ along $\ket{3z_{\alpha}^{2}-r^{2}}$ 
(where $\alpha=\{x,y,z\}$ is the bond orientation). 
This derivation, the details of which are to lengthy to be presented 
here, leads to the following Hamiltonian in the basis of $e_{g}$ 
orbitals 
$\{\ket{z}=\ket{3z^{2}-r^{2}}, \ket{\bar{z}}=\ket{x^{2}-y^{2}}\}$:
\begin{multline}
  \label{eq:trans3d}
  \mathcal{H}_{t} = -t [\sum_{\mean{ij} \| z} d_{iz}^{\dag} d_{jz}^{} +
  \frac{1}{4}\sum_{\mean{ij} \| x} (d_{iz}^{\dag} -\sqrt{3} d_{i\bar{z}}^{\dag})
  (d_{jz}^{} -\sqrt{3} d_{j\bar{z}}^{}) \\
  +\frac{1}{4}\sum_{\mean{ij} \| y} (d_{iz}^{\dag} +\sqrt{3} d_{i\bar{z}}^{\dag})
  (d_{jz}^{} +\sqrt{3} d_{j\bar{z}}^{})] + \mathrm{H.c.},
\end{multline}
with orbital exchange interaction
\begin{equation}
  \label{eq:Hj}
  \mathcal{H}_{J} = \frac{J}{2} \sum_{\mean{ij}} [
  T_{i}^{z}T_{j}^{z} + 3 T_{i}^{x} T_{j}^{x}
  \mp \sqrt{3} (T_{i}^{x}T_{j}^{z}+T_{i}^{z}T_{j}^{x}) ],
\end{equation}
where the operators $T^{\alpha}$ are analogous to regular spin operators 
(\emph{i.e.}, 1/2 times the respective Pauli operator), only acting in 
the $e_{g}$ orbital space spanned by the $\{\ket{z},\ket{\bar{z}}\}$ basis.

Henceforth we shall neglect the orbital fluctuations and only keep the 
Ising part of the exchange Hamiltonian. The reason is that the 
fluctuations have a smaller amplitude than the leading term of the 
Hamiltonian \cite{Cin10}, while the kinetic energy \eqref{eq:trans3d} 
does not conserve the orbital flavor and so is the source of orbital 
excitations on a much bigger scale. Furthermore, we will be performing an 
expansion around a N\'eel-type ground state, which is inconsistent 
with keeping the orbital fluctuations.

Since the Ising part of the Hamiltonian is the leading 
$3T_{i}^{x} T_{j}^{x}$ term, the basis states are the eigenstates of 
those operators, \emph{i.e.}, $\ket{\pm}=(\ket{z}\pm\ket{\bar{z}})/\sqrt{2}$, 
and so the Hamiltonian needs to be transformed accordingly. Taking into 
account that the exchange coupling constant $J$ is positive, 
we can ascertain that 
the orbital ground state exhibits an alternating orbital (AO) order.

At this point it is useful to decouple the orbital degree of freedom 
from the fermionic operators by means of slave boson $a^{(\dag)}$ 
formalism:
\begin{equation}
  \label{eq:slave}
  d_{i0}^{\dag} = f_{i}^{\dag}, \quad d_{i1}^{\dag} = f_{i}^{\dag} a_{i}^{},
\end{equation}
where the $\{0,1\}$ indices denote the ground or excited orbital state, 
\emph{i.e.}, $\ket{+}$ or $\ket{-}$ depending on the sublattice. After 
performing those transformations the resulting Hamiltonian takes the 
final form
\begin{subequations}
  \label{eq:transfH}
  \begin{align}
    \label{eq:transfHJ}
    \mathcal{H}_{J} &= \frac{3}{8}J\sum_{\mean{ij}} (1 - \sigma_{i}^{z}\sigma_{j}^{z}),\\
    \label{eq:transfT}
    \mathcal{T} &= \frac{t}{4}\sum_{\mean{ij}} (f_{i}^{\dag}f_{j}^{} + \mathrm{H.c.}) =
    \sum_{\mathbf{k}} \epsilon_{\mathbf{k}}^{} f_{\mathbf{k}}^{\dag} f_{\mathbf{k}}^{},\\
    \label{eq:transfV}
    \mathcal{V} &= \frac{t}{4} \sum_{i,\delta} \left[
      (2+\sqrt{3} e^{i\pi_{y}\cdot\delta}
      e^{i \mathbf{Q}\cdot\mathbf{R}_{i}}) a_{i}^{\dag} +\right.\nonumber\\
    &\left.+(2-\sqrt{3} e^{i\pi_{y}\cdot\delta}
      e^{i \mathbf{Q}\cdot\mathbf{R}_{i}})
      a_{i+\delta}^{} + a_{i}^{\dag} a_{i+\delta}^{} \right]
    f_{i+\delta}^{\dag} f_{i}^{},
  \end{align}
\end{subequations}
where $\epsilon_{\mathbf{k}}^{} = t\gamma_{\mathbf{k}}^{}$ is the energy 
of a free particle, with 
$\gamma_{\mathbf{k}}^{}=(1/z)\sum_{\delta} e^{i\mathbf{k}\cdot\delta}$. 
For simplicity, \eqref{eq:transfHJ} has been transformed from AO to 
ferro-orbital state by a rotation on one sublattice, which changes the 
overall sign of the interaction, hence the $-3/8$ factor in front of $J$. 
Note the constant added to the Hamiltonian to put the ground state 
energy at zero to simplify the calculations. The interaction 
$\mathcal{V}$ comes from the kinetic Hamiltonian and is a consequence 
of the orbital flavor non-conservation of the model. The phase factors 
$\pi_{y} = (0,\pi)$, $\mathbf{Q}=(\pi,\pi)$ serve to incorporate the 
model's dependence on direction and delta is the vector pointing to 
a site's nearest neighbors.

The variational approximation consists in a series of Dyson expansions,
\begin{equation}
  \label{eq:dyson}
  \mathcal{G}(\omega) = \mathcal{G}_{0}(\omega)
  +\mathcal{G}(\omega)\mathcal{V}\mathcal{G}_{0}(\omega),
\end{equation}
to generate the equations of motion for the Green's function, where 
$\mathcal{H}_{0} = \mathcal{T}+\mathcal{H}_{J}$ corresponds to 
$\mathcal{G}_{0}(\omega)$ and $\mathcal{V}$ is given by 
Eq.~\eqref{eq:transfV}. Let us define the Green's function as
$G(\mathbf{k},\omega)\equiv\bra{\mathbf{k}}\mathcal{G}(\omega)\ket{\mathbf{k}}$,
where $\mathcal{G}(\omega)=(\omega+i\eta-\mathcal{H})^{-1}$ is the 
resolvent and
\begin{equation}
\ket{\mathbf{k}}\equiv f_{\mathbf{k}}^{\dag}\ket{0}=
\frac{1}{\sqrt{N}}\sum_{i}e^{i\mathbf{k}\cdot\mathbf{R}_i}f_{i}^{\dag}\ket{0},
\label{eq:kstate}
\end{equation}
is the free electron Bloch state. The core idea underlying the 
variational approximation is that the energy cost of an orbiton 
creation is proportional to $J$, hence for large $J$ only a small 
number of orbitons can be created \cite{Tro13}.

Since $G_{0}(\mathbf{k},\omega)$ is known and diagonal in $\mathbf{k}$, 
the key part of the Dyson expansion is evaluating 
$\mathcal{V}\ket{\mathbf{k}}$, which is done in real space, leading to
\begin{equation}
  \label{eq:green}
  \begin{split}
&G(\mathbf{k},\omega) = \left[ 1-\frac{t}{2}\sum_{\delta}
F_{1}(\mathbf{k},\omega,\delta)+\right.\\
&\left.-\frac{\sqrt{3}t}{4}\sum_{\delta} \bar{F}_{1}(\mathbf{k},\omega,\delta)
e^{i\mathbf{\pi}_y\cdot\mathbf{\delta}} \right] G_{0}(\mathbf{k},\omega-4J'),
  \end{split}
\end{equation}
where $J'=\frac{3}{8}J$ and the generalized Green's functions
\begin{align}
  \label{eq:f1}
F_{1}(\mathbf{k},\omega,\delta) &=
\bra{\mathbf{k}} \mathcal{G}(\omega)
\frac{1}{\sqrt{N}} \sum_{i} e^{i\mathbf{k}\cdot\mathbf{R}_{i}}
f_{i+\delta}^{\dag} a_{i}^{\dag} \ket{0},\\
\label{eq:f1b}
\bar{F}_{1}(\mathbf{k},\omega,\delta) &=
\bra{\mathbf{k}} \mathcal{G}(\omega)
\frac{1}{\sqrt{N}} \sum_{i} e^{i(\mathbf{k+Q})\cdot\mathbf{R}_{i}}
f_{i+\delta}^{\dag} a_{i}^{\dag} \ket{0}.
\end{align}

These functions are unknown and need to be calculated by further 
Dyson expansions which, after applying $\mathcal{V}$ to the 
$f_{i+\delta}^{\dag} a_{i}^{\dag} \ket{0}$ state, generate other Green's 
functions, such as $G(*,*), F_{1}(*,*,*)$ and the 2-orbiton functions
\begin{equation}
  \label{eq:f2}
  F_{2}(\mathbf{k},\omega,\delta,\epsilon) =
\bra{\mathbf{k}} \mathcal{G}(\omega)
\frac{1}{\sqrt{N}} \sum_{i} e^{i\mathbf{k}\cdot\mathbf{R}_{i}}
f_{i+\delta+\epsilon}^{\dag} a_{i+\delta}^{\dag} a_{i}^{\dag} \ket{0},
\end{equation}
which also need to be expanded 
further. This process 
could be continued indefinitely, so at some point the equations have 
to be cut by disallowing the creation of any further orbitons in the 
system, hence it is controlled by the number of orbital excitations.

Once the system has more than one orbiton, there are numerous ways to 
de-excite it, namely at each step an orbiton can be removed from either 
end of the string. In particular, destroying an orbiton other than the 
one created last is a process analogous to the crossing-diagrams excluded 
in SCBA. Here we try to establish the importance of such processes by 
comparing the Green's functions which include or exclude them in the 
2-orbiton regime.

After the first orbiton is added, certain constraints 
have to be imposed on the electron's movement, namely: 
($i$) the electron cannot occupy the same site as the orbiton, and  
($ii$) in the case where the electron is on a site adjacent to the 
orbiton the $\mathcal{H}_{J}$ energy increase is $10J'$, compared to 
the regular energy $12J'$ when the particles are far apart. 
Because of this, the translational invariance is broken, so that 
$\mathbf{k}$ is no longer a good quantum number. Therefore, at higher 
levels of the expansion one has to calculate real space Green's 
functions while including the above constraints, which is added as a term to 
the Hamiltonian $\mathcal{H}_{0}$ to cancel the corresponding processes:
\begin{equation}
  \label{eq:vi}
  \mathcal{V}_{1} = -\frac{t}{4} \sum_{\epsilon}
  (f_{i}^{\dag} f_{i+\epsilon}^{} + \mathrm{H.c.})
  -2J'\sum_{\epsilon} n_{i+\epsilon},
\end{equation}
where $i$ is the location of the orbiton and $n_{i+\epsilon}$ is the 
electron number operator. The constrained non-interacting Green's 
function is then calculated from the non-constrained one similarly, 
by Dyson expansion
\begin{equation}
\mathcal{G}_{1}(\omega) = [1+\mathcal{G}_{1}(\omega)\mathcal{V}_{1}]
\mathcal{G}_{0}(\omega),
\end{equation}
which leads to a matrix equation, describing propagations between the 
orbiton's neighboring sites
\begin{equation}
  \label{eq:gimatrix}
\mathbb{G}_{1}^{\gamma\delta} = \mathbb{G}_{0}^{\gamma\epsilon}
\left[\mathbb{I}^{\delta\epsilon} +\tfrac{t}{4}\mathbb{G}_{0}^{\delta 0}
+2J'\mathbb{G}_{0}^{\delta\epsilon}\right]^{-1},
\end{equation}
where the Greek indices denote the orbiton's neighboring sites, 
so the matrix element 
$\mathbb{G}_{1}^{\gamma\delta} = G_{1}(\gamma,\delta,\omega)$ is the 
constrained Green's function describing the 
$\ket{i+\delta}\rightarrow\ket{i+\gamma}$ propagation in real space. 
A similar equation is found for the case of two or more orbitons, only 
the indices run over all the neighboring sites of the orbiton string.

Once the equations of motion for the Green's function are generated and 
cut at the desired level (in this work at two orbitons), one is left 
with a set of equations for various Green's functions. In principle the 
system can be solved for all of them, but usually we are only interested 
in the normal Green's function $G(k,\omega)$. However, what is usually 
plotted is the normalized spectral function
\begin{equation}
  \label{eq:spec}
  A(\mathbf{k},\omega)=-\frac{1}{\pi}\,\Im[G(\mathbf{k},\omega)],
\end{equation}
which has the interpretation of the quasiparticle density of states. 
Furthermore, here we plot $\tanh[A(k,\omega)]$, which 
amplifies 
the low amplitude part of the spectra, while treating the large 
amplitudes almost uniformly by mapping them into values close to 1.

\section{Results and Discussion}
\label{sec:res}

\begin{figure}[t!]
  \centering
  \includegraphics[width=\columnwidth]{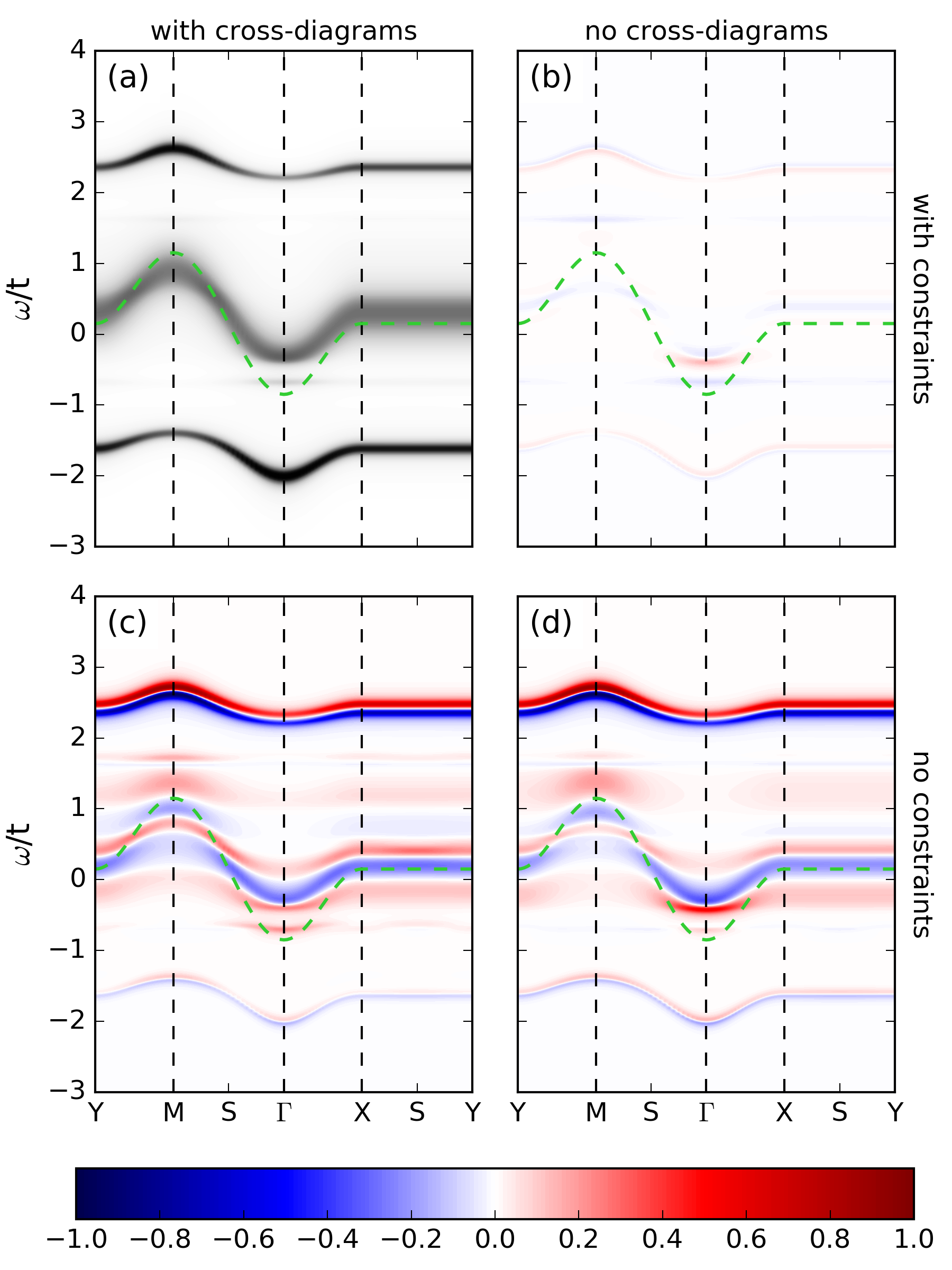}
\caption{Spectral functions $A(\mathbf{k},\omega)$ for $J/t=0.1$: 
(a) the full function, including cross-diagrams and constraints, 
(b) difference for the case without cross-diagrams, 
(c) difference for the case without constraints, 
(d) difference for the case without both effects. 
The dashed green line indicates the free electron dispersion 
$\omega=\epsilon_{k}+4J'$ for reference. 
The colorbar refers only to panels (b)--(d). Note the $\tanh$-scale.}
  \label{fig:spec}
\end{figure}

In Fig. \ref{fig:spec} the calculated spectral functions are shown 
in a nonlinear $\tanh$-scale to emphasize the low amplitude features 
of the spectral functions. Panel (a) shows the full Green's function, 
including the crossing diagrams and the translational constraints,
while panels (b)--(d) focus on the difference functions when constaints are 
neglected, see below. 

The huge advantage of the VA is that it is an 
analytical method with precise control of the states spanning the 
Hilbert space. When performing the expansion and evaluating the 
interaction, one can easily include or omit processes according to 
their importance or likelihood of occurrence. For instance, if the 
system, after creating multiple bosons, starts removing them in an 
order reverse to the order of creation, then it is a non-crossing 
process, because the bosonic lines of its Feynman diagram can never 
cross. Any other sequence of boson removals leads to crossing diagrams, 
and the relative number of such processes is the bigger the more 
bosons there are in the system. However, their importance can be hard 
to ascertain as it mostly depends on the interaction vertex and the 
boson energy. Knowing the significance of the crossing-diagrams is very 
important for using methods like the SCBA, since it is an approximation 
that by its very nature includes only non-crossing diagrams, while 
adding other processes can be very tricky. However, since VA makes 
it easy to turn those processes on or off, it is a good method to check 
their role, even if only within a low expansion order.

On the other hand, SCBA, being a Fourier space expansion, requires full
translational invariance. However, as already explained, this is broken
once there are bosons in the system. Therefore, SCBA simply ignores
that, assuming that for a big system with a small number of bosons the
lattice is almost fully translational invariant. However, polaronic
physics is strongly local, with all the interactions happening in the
vicinity of the boson. Therefore, in general the translational
constraints are expected to play a crucial role. Unfortunately,
SCBA cannot include those effects at all, while VA does it exactly
and fully.
In this paper we use the VA \cite{Ber11} to examine the importance of
the two effects described above. To do so, we calculate Green's
functions including both of the effects, excluding either of them,
or excluding both.



Panel (b) shows the difference function for the case without 
cross-diagrams, but including translational constraints. Somewhat 
surprisingly, we see that at the lowest order of expansion the 
cross-diagrams play a very small, almost negligible role, with 
maximal values of amplitude change at around 5\%. The qualitative 
change of the spectrum is also very subtle, with only a tiny 
transfer of weight at $\Gamma$ and $M$ points and a small reduction 
of bandwidth (indicated by a pair of parallel red-blue
lines in the function, which mean that the reference 
maximum of panel (a) has to move away from the blue line and
towards the red line).

On the other hand, the effect of the translational constraints is very 
strong. Panel (c) shows that neglecting this effect causes the QPs to 
gain additional energy, by shifting the whole spectrum upwards. This 
is especially dramatic for the excited state in the upper part of the 
spectrum where the shift is around $0.13t$ and the amplitude is very 
big, while in the ground state at the bottom the effect is somewhat 
smaller, with a shift of $0.06t$ and the amplitude change of around 
3\%. This however is still bigger than the effect of cross-diagrams. 
In the incoherent part of the spectrum in the middle, the influence 
of the constraints is quite strong but qualitatively complicated. 
The exclusion of constraints seems to narrow the width of the 
pseudo-band visible in the middle of panel (a) on one hand, and tend 
to split the band into two around $\omega=0$ on the other, but not 
enough to separate them completely. This has the effect that although 
the spectrum in the middle becomes more coherent, it appears even 
less so because the various bands blend together. This in principle 
is in accordance with SCBA, which shows a broad incoherent continuum, 
with a barely discernible ladder of low amplitude states. However, 
this effect does not seem to account for the whole difference, since 
the two results still differ quite substantially.

Finally, panel (d) shows the results when both cross-diagrams and 
constraints are turned off. Since the two effects are completely 
independent and do not interfere with each other, it is no surprise 
that their combined effect does not differ much from those effects 
treated separately. In particular, since the cross-diagram effects 
are so small, it is clear that the results in panel (d) are nearly 
identical to those in panel (c). A close inspection might reveal 
that some of the features are more pronounced, especially where the 
two effects would combine positively, such as in the ground state 
or in the incoherent part around the $\Gamma$ point. However, 
qualitatively the picture remains mostly the same.

In conclusion, we have shown, using a highly accurate and primarily 
analytical method, the effects of two most important approximations 
employed by the SCBA, which is a standard method widely used in 
polaronic physics. We have shown that within our $e_{g}$ orbital model 
the effects of cross-diagrams is very small and can mostly be neglected. 
However, due to the highly local nature of the polaronic QPs, the 
translational constraint effects are very important and cannot 
conceivably be neglected. Furthermore, they can to some extent explain 
the difference between SCBA and VA, although not fully. Since there is 
no way to include those constraints in SCBA, that method should be used 
with caution.

\begin{acknowledgments}

We thank Krzysztof Wohlfeld for insightful discussions. 
We kindly acknowledge support by UBC Quantum Matter Institute, 
by Natural Sciences and Engineering Research Council of Canada (NSERC), 
and by Narodowe Centrum Nauki (NCN) under Projects 
No.~2012/04/A/ST3/00331 and 2015/16/T/ST3/00503.

\end{acknowledgments}

\end{document}